# THE WIDE-FIELD NEAR INFRARED DATA: OPTIMAL PHOTOMETRY IN CROWDED FIELDS

R.R. King and Tim Naylor[*]
School of Physics, Stocker Road, University of Exeter, Exeter EX4 4QL, UK

Patrick S. Broos, Konstantin V. Getman and Eric D. Feigelson
Department of Astronomy & Astrophysics, Pennsylvania State University, 525 Davey Lab, University Park, PA 16802
*Draft version September 17, 2013*

## ABSTRACT

We present $JHK$ infrared data from the UK Infrared Telescope for a subset of the regions of the MYStIX (Massive Young Star-Forming Complex Study in Infrared and X-ray) survey. Some of the data were obtained specifically for the MYStIX project, and some as part of the UKIRT Infrared Deep Sky Survey's Galactic Plane Survey. In most of these fields crowding is a significant issue for aperture photometry, and so we have re-extracted the photometry from the processed images using an optimal extraction technique, and we describe how we adapt the optimal technique to mitigate the effects of crowding.

*Keywords:* methods: data analysis − techniques: photometric − stars: pre-main sequence − infrared: stars − stars: formation − open clusters and associations: general

## 1. INTRODUCTION

Study of the pre-main sequence population of massive star forming regions is impeded by obscuration in molecular clouds, nebulosity in H II regions, and contamination by foreground and background Galactic field stars. The Massive Young Star-Forming Complex Study in Infrared and X-rays (MYStIX; Feigelson et al. 2013) seeks to treat these difficulties from a multiwavelength perspective. For 20 massive star-forming regions, X-ray images from the *Chandra X-ray Observatory*, near-infrared images from the United Kingdom InfraRed Telescope (UKIRT), mid-infrared images from the *Spitzer Space Telescope*, and published lists of spectroscopically identified OB stars are combined to produce samples of probable young members. Each waveband of the observational database for MYStIX provides information about different emission mechanisms of the young stellar objects.

The primary role of the $JHK$ infrared (IR) data is to reveal the stellar photospheres, free from contamination (at least at $J$ and $H$) by emission from circumstellar disks. Although optical data are often used for this purpose, many of the MYStIX clusters are sufficiently reddened that near-IR data are the only practical route. An obvious source of such data is the 2MASS survey (Skrutskie et al. 2006), but in the 11 fields listed in Table 1 we found that 2MASS lacked detections in one or more bands (at a signal-to-noise greater than 10) for roughly 30 percent of the X-ray sources, and 50 percent of the sample identified as MYStIX Probable Complex Members (MPCMs; see Broos et al. 2013). In anticipation of these problems, we acquired new deep $JHK$ data for six of the MYStIX targets using the UK Infrared Telescope (UKIRT), and added to these data for another five which were present in the UKIRT Infrared Deep Sky Survey (UKIDSS; Lawrence et al. 2007).

In this paper we describe the data acquisition and reduction, and then in Naylor et al. (2013) we describe cross matching the resulting catalogs with the *Chandra* data described in Kuhn et al. (2013a) and Townsley & Broos (2013) to provide IR counterparts. The analysis which follows in subsequent papers is a multi-wavelength approach in which the contribution of each band is not explicitly made clear, but the contribution of the $JHK$ data is broadly as follows. An X-ray source with a faint UKIRT counterpart is likely to be classified as a background extragalactic source by the Bayesian classifier described in Broos et al. (2013). Stars with brighter $J$-band magnitudes are likely to be classified as field stars or MPCMs. Although $JHK$ data have historically been used to identify sources with disks (e.g., Mendoza V. 1966; Rydgren et al. 1976; Kenyon & Hartmann 1987; Strom et al. 1989), it has long been recognized that longer wavelength data are required to construct a complete sample of stars with disks (e.g. Haisch et al. 2000). Thus we never use the $JHK$ data alone to search for disks. Its combination with our *Spitzer* data presented in Kuhn et al. (2013b) makes the powerful tool for the identification of young stars with disks we describe in Povich et al. (2013). Finally, once a young star is reliably identified, the $J$-band data combined with the measurement of the extinction which relies on the *Spitzer* data, gives a measure of the photospheric emission.

In this paper we first describe the NIR (near infrared) data acquisition (Section 2) and then the data reduction. As we discuss in Section 3 the standard UKIDSS data products are not well suited to the crowded fields of the Galactic plane, and so we present an adaption of the optimal extraction in Sections 5 and 6. The data products are discussed in Sections 7 to 9.

## 2. DATA ACQUISITION

The data were obtained using WFCAM (Casali et al. 2007), the infrared wide-field camera on UKIRT in Hawaii. Roughly half the fields were observed as part of the Galactic Plane Survey (GPS; Lucas et al. 2008) component of the UKIDSS with the remainder being obtained in Director's Discretionary Time (DDT) using identical observing procedures.

---
[*] timn@astro.ex.ac.uk

We aimed to present data not just for the star-forming regions themselves, but also for a considerable area around them, so that statistical studies of the field populations could be undertaken. Thus for the DDT data we observed a rectangular area 0.9 by 0.9 degrees, with the target at its center. For the survey data, we had to select pointings which covered each star-forming complex, and a generous surrounding area. The WFCAM focal plane contains four arrays with large (almost one array width) gaps in between. This allows for rapid survey work, but also means that sometimes, to obtain one crucial array, another four which are not connected to them had also to be reduced (since this helped our photometric solution – see Section 6). We have included these data in our final catalogue, but it can result in the geometry of our data boundary being complex at the periphery of our chosen field. Hence in Table 1 we give the central co-ordinates and maximum extent of each field, even if the field is not completely filled.

At each position, there are eight exposures of 10s in $J$ and $H$ and 5s in $K$. The depth reached by these exposures can depend critically on the crowding, and so in Table 1 we present the modal $K$-band magnitude in each field for all objects both unflagged (see Section 7.1) and with a signal-to-noise greater than 10 in $K$, as an indication of the limiting magnitude reached. In uncrowded fields the number of objects detected per unit magnitude will fall to a tenth of this peak by 0.5 mags fainter, whilst for crowded fields that "roll off" can take 1.5 magnitudes.

## 3. DATA REDUCTION - OVERVIEW

It was recognized in the planning stages of UKIDSS that the standard pipeline, based on aperture photometry, might not be able to extract the best possible photometry in crowded fields, and that this was a particular concern for the Galactic plane (see Section 6.2 of Lawrence et al. 2007). Therefore for the MYStIX project we use the optimal extraction of Naylor (1998). This uses the idea that if the stellar profile is understood, the fraction of the stellar flux in each pixel of the stellar image can be calculated. If one divides the (background subtracted) counts in a given pixel by that fraction, one obtains a measurement of the total stellar flux. Each pixel in the stellar profile therefore provides an independent measure of the brightness of the star, and combining these measures using a weighting which maximizes the signal-to-noise of the final flux determination, yields the optimal measure (in terms of signal-to-noise) of the stellar brightness. Although it does not explicitly separate the flux of overlapping stellar images, the fact that the measurement of the flux is heavily weighted to the central pixels of the stellar image means that it performs better than aperture photometry in crowded fields. For isolated stars optimal extraction is mathematically equivalent to profile fitting, but the optimal extraction is more suited to an unsupervised pipeline as it is simpler and less computationally intensive. In addition, it provides a better estimate of the uncertainty in the flux for reasons described in Naylor (1998).

Our data reduction strategy was to use the processed images from the UKIDSS pipeline (see Section 4) and pass them through the optimal extraction software (Section 5). This results in raw magnitudes, which we then tied to both the UKIDSS photometric system and the astrometric reference frame using 2MASS (Section 6).

## 4. IMAGE PROCESSING AND SOURCE DETECTION

The individual images processed through the standard UKIDSS pipeline (Dye et al. 2006, Irwin et al. in prep) were recovered from the WFCAM Science Archive (Hambly et al. 2008). This provides images which are largely free of the instrumental signature. It also yields a confidence map, in which pixels with a low (or zero) sensitivity, or ones with unpredictable levels are given a low confidence. We used this image to flag pixels which we then ignored through the remainder of the reduction process.

A significant difference between the UKIDSS and our reduction is that we carry out our photometry on the individual images (not their sums). To achieve this, however, we still need a deep image to detect the sources in, which we obtained by combining the $K$-band images for a given pointing (by simple integer pixel shifts) to produce a deep image. The source detection algorithm (Naylor et al. 2002; Jeffries et al. 2004) then yielded a list of objects (and of bright stars with could be used to model the mean stellar profile) on which photometry is carried out in all the available images in all available bands. The procedure results in reliable upper limits for stars which are detected in $K$, but not in one or more of the bluer bands.

## 5. OPTIMAL EXTRACTION

A detailed description of optimal extraction and its implementation for constructing color-magnitude diagrams is given in Naylor (1998) and Naylor et al. (2002). Those papers concentrated on using the extraction for stars which were relatively uncrowded, and so the extraction parameters were optimized for such fields. Once the field is crowded, as we shall show later, more accurate photometry can be obtained by sacrificing formal signal-to-noise to decrease the effects of spill-over. Hence in this paper we will concentrate on the changes which have to be made to optimize the technique for crowded fields.

### 5.1. *The profile clipping radius*

#### 5.1.1. *Signal-to-noise in the well-sampled case*

Optimal photometry integrates the flux after applying a weight mask that tends to zero at large radii. Practically, this mask is set to zero beyond some "clipping radius". The effect of spill-over from other stars can be mitigated by reducing this radius, but with some loss of signal-to-noise. That loss in signal-to-noise can be calculated for a given clipping radius by integrating the profile in Equation (12) of Naylor (1998) out to that radius, and this is shown in Figure 1. In Naylor (1998) the clipping radius was set to twice the FWHM seeing to maximize the signal-to-noise. Figure 1 makes clear that this choice was very conservative. To avoid contamination from other stars in crowded fields that radius could easily be halved and have little impact on the signal-to-noise.

#### 5.1.2. *Signal-to-noise in the poorly-sampled case*

Table 1
Summary of observations.

| Name | Central Position | | Extent (deg) | | Number of arrays | Number of Objects | Modal K-mag | Objects per arcmin$^2$ |
| --- | --- | --- | --- | --- | --- | --- | --- | --- |
| | RA (J2000) | Dec (J2000) | RA | Dec | | | | |
| (1) | (2) | (3) | (4) | (5) | (6) | (7) | (8) | (9) |
| NGC 2264 | 06 40 56.1 | +09 40 21.89 | 0.89 | 0.89 | 16 | 42587 | 17.5 | 15 |
| Rosette Nebula | 06 32 30.7 | +04 26 05.05 | 2.35 | 1.55 | 64 | 159780 | 17.2 | 14 |
| Lagoon Nebula | 18 03 56.1 | −24 11 29.42 | 0.89 | 0.89 | 16 | 530072 | 16.0 | 184 |
| NGC 2362 | 07 18 46.4 | −24 57 42.66 | 0.89 | 0.89 | 16 | 56289 | 17.3 | 20 |
| DR 21 | 20 39 16.8 | +42 18 23.80 | 1.80 | 1.77 | 56 | 542796 | 17.4 | 54 |
| NGC 6334 | 17 20 24.8 | −35 51 55.78 | 0.89 | 0.89 | 16 | 655046 | 17.0 | 227 |
| NGC 6357 | 17 25 32.8 | −34 16 34.13 | 0.89 | 0.89 | 16 | 689910 | 16.9 | 240 |
| Eagle Nebula | 18 20 21.5 | −13 48 52.53 | 1.80 | 1.77 | 64 | 3280391 | 16.7 | 285 |
| M 17 | 18 20 06.7 | −16 08 27.99 | 1.24 | 1.14 | 32 | 1092373 | 16.4 | 190 |
| Trifid Nebula | 18 02 26.0 | −22 52 15.22 | 1.55 | 1.77 | 40 | 1668056 | 16.0 | 232 |
| NGC 1893 | 05 22 52.5 | +33 28 42.80 | 0.89 | 0.89 | 16 | 54889 | 17.8 | 19 |

**Note**. — Col. 12 gives the number of unflagged objects with a signal-to-noise greater than 10 in $K$.

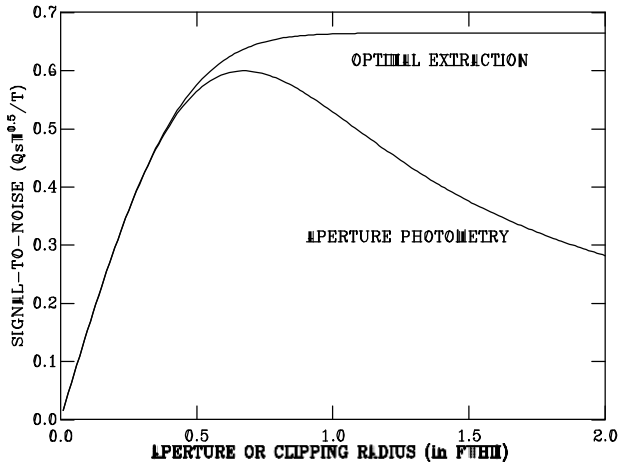

**Figure 1.** The signal to noise as a function of clipping radius for the optimal extraction, and as a function of aperture radius for aperture photometry. The units are signal divided by the square root of the variance per unit area of FWHM seeing.

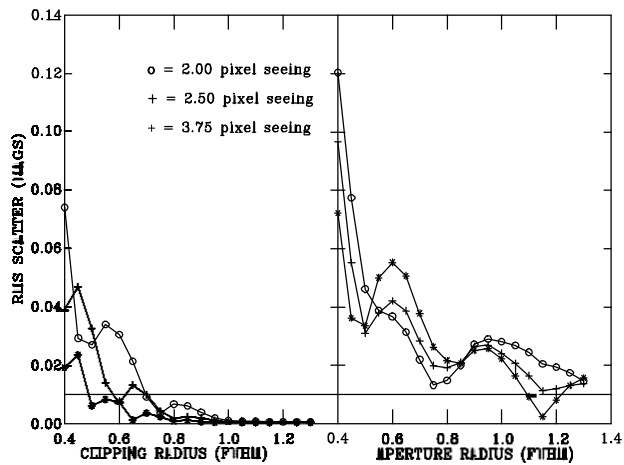

**Figure 2.** The variation in magnitude due to resampling noise for optimal extraction as a function of clipping radius (left) and for aperture photometry as a function of aperture radius (right).

Figure 1 assumes the pixels are small compared with the mask size. However, in WFCAM data the seeing is typically only 2.5 pixels, and so there is a resampling noise, due to interpolating the pixel values onto the mask (see Section 2.2 of Naylor 1998). To assess the impact of this resampling noise on our photometry we simulated 25 stars of identical magnitude sampled onto the pixel grid so the fractional pixel positions of the stars were evenly spaced in $x$ and $y$ with a separation of 0.2 pixels. The stellar profiles were Gaussians, numerically integrated over the area of each pixel, with equal total fluxes. We then passed these simulated images through the optimal extraction. We dealt with pixels at the aperture edge in the way described in Da Costa (1992). Whilst this commonly used scheme is not beyond reproach, as we shall see later it reduces the resampling noise for optimal photometry to a negligible level.

The left panel of Figure 2 shows the measured resampling noise for values of the seeing typical for our data, as a function of the clipping radius. As might be expected the most significant effects are seen for the best seeing, which of course is the data we would normally expect to have the best signal-to-noise. If we choose a clipping radius of 0.8 times the FWHM seeing, it is clear the RMS contribution from resampling noise will always remain below a percent. We also experimented with defining the clipping radius in terms of pixels rather than the seeing, but found that the noise relationships for different seeings were not as tightly clustered, so one could not not use a simple criterion for the clipping radius.

The right panel of Figure 2 shows the resampling noise for aperture photometry; it is immediately apparent that resampling noise is much worse than for optimal photometry. This result is explained by the low weight that pixels near the profile edge have in optimal photometry, compared with the full weight for aperture work.

### 5.2. Background box size

Correctly determining the night sky background level is key for $JHK$ photometry, where the background flux is typically far greater than that of the star. Virtually all packages determine the background as the modal value in an area surrounding the star, thus rejecting any contributions from stars which lie within that area. We determine the mode by fitting the distribution of pixel values with a skewed Gaussian. Using a large area results in the sky being determined from a large number of pixels, and hence with a high degree of precision. However, whilst this may result in an accurate determination of the sky area as a whole, if the background is structured (as our backgrounds inside nebular HII regions often are) it may result in a value which does not reflect the background at the position of the star. Clearly therefore there is a trade-off to be made.

A useful quantity for assessing the best size of sky box to be used is $3i$, the uncertainty in the stellar flux from the photons within the clipping radius, divided by the uncertainty due to the sky subtraction. We can adapt Equation 16 of Naylor (1998) to show that this is given by

$$N_s > \frac{3i^2 s^2 \pi}{2 \ln 2},\qquad(1)$$

where $N_s$ is the number of pixels used to estimate the mean sky level and $s$ is the seeing in pixels. Naylor (1998) aimed for an $3i$ of 14, which results in sky boxes of side $21s$. If instead we state that the sky estimate is to increase the final uncertainty by a factor of no more than one percent, this corresponds to $3i = 7$. So, for example, a star with a photon uncertainty of 10 percent, would have an additional uncertainty from the sky of 1.4 percent, which when added in quadrature would give a final uncertainty of 10.1 percent. This has the advantage of reducing our sky boxes to a side of $10s$ pixels, or typically $10''$.

### 5.3. Final extraction parameter choices

Summarizing this section, we can see that if we wish to reduce the weight-map clipping radius in order to decrease the contamination from nearby stars, our primary concern must be that the resampling noise does not become too large. We have therefore chosen a clipping radius of 0.8 times the FWHM seeing, which is considerably smaller than the value of twice the FWHM we have used in previous reductions. This value of the clipping radius always results in a resampling noise of less than one percent. Section 5.1.1 shows that the resulting loss in signal-to-noise due to using a not-quite-optimal weight map is also less than a percent. Similarly, we have shown that we can reduce the sky box size to have a side of 10 times the FWHM seeing, with little loss in signal-to-noise.

### 5.4. Combining the photometry from each image

Prior to photometric calibration we combined the measurements from each individual image using a weighted mean, to produce instrumental $J$, $H$ and $K$ magnitudes, in addition to $J-K$ and $H-K$ colors. This provides an additional opportunity to test the robustness of the uncertainties using the scatter about the weighted mean. At bright magnitudes the scatter between measurements is independent of magnitude and is about 2 percent, and hence this is added in quadrature to the photon uncertainties for each star, as explained in Naylor et al. (2002). For faint stars the scatter about the means implies the uncertainties from single images are under-estimated by a factor of approximately 1.2. For faint stars the estimate of the uncertainty from single images is driven by the measurement of the noise in the sky photons which originates from the fit to the histogram of sky counts described in Section 5.2 (see also Section 3 of Naylor 1998), which relies on the assumption that the pixel values are uncorrelated. In fact, stray capacitance between the pixels results in a decrease of the variance by approximately 20 percent (Dye et al. 2006), explaining a significant fraction of the factor of 1.2.

### 6. ASTROMETRIC AND PHOTOMETRIC CALIBRATION

We used positions from the 2MASS catalogue (Skrutskie et al. 2006) to obtain an astrometric solution that had a typical RMS accuracy of $0.09''$ in either RA or Dec, implying that 68 percent of all our positions lay within a radius of $0.13''$ of their 2MASS counterparts.

The aim of our photometric calibration was to yield colors and magnitudes whose photometric system matched that of the standard UKIDSS reduction as closely as possible, by following the procedure described by Hodgkin et al. (2009), appropriately adapted for optimal photometry. Normally a photometric reduction with simple apertures uses an aperture which excludes much of the flux in order to obtain a good signal-to-noise for the faint stars. In an analogous way an optimal extraction will not match the full flux from the star unless the profile used to extract the flux exactly matches the true stellar profile. In both cases we can expect the correction to be a smooth function of position on the array for any one observation, since it is simply a function of the stellar profile which will vary with both the seeing (which we can expect to be uniform over an array at any particular time) and the telescope optics. We therefore compare our optimally extracted magnitudes with those of 2MASS, which allows us to simultaneously correct for the profile mis-match (the "profile correction" of Naylor et al. 2002) and photometrically calibrate our data.

If the UKIDSS system responses were identical to those of 2MASS, this would be a straightforward process, but there are small differences, which mean that stars will not necessarily have the same magnitude in the 2MASS and UKIDSS systems. As discussed, for example, in Bell et al. (2012) no set of photometric transforms can overcome this in the general case where the objects are not main-sequence stars. This led Hodgkin et al. (2009) to define the UKIDSS system as the natural system, with the zero point fixed by the stipulation that main-sequence stars of color zero have identical colors and magnitudes in both the UKIDSS and 2MASS systems. Given that there are many 2MASS stars in our fields of view, this in principle provides us with a method of calibrating the data, but in practice there are never enough stars around zero color to make this practical. So we have to compromise and simply use stars which are not very red, matching our selection criteria as closely as possible to those of Hodgkin et al. (2009) to ensure comparable data. In addition, like Hodgkin et al. (2009) we transform the 2MASS data to the UKIDSS system before making the comparison, since most stars will be close to main-sequence spectral energy distributions.

To obtain our sample of blue stars we selected all the stars with (i) good quality photometry in our data and 2MASS (ii) an uncertainty in the difference in magnitude between our reduction and 2MASS of less than 0.04 mags and (iii) $J-K<1$. If there were less than 25 such stars on the array, we chose the 25 bluest objects that satisfied conditions (i) and (ii). We created the profile correction by fitting the magnitude differences as a function of po-sition on the array with a two dimensional polynomial of order two in both directions.

**Table 2**
Column headings for the UKIRT catalog.

| Label | Units | Description |
|---|---|---|
| MYSTIX_SFR | | MSFR name |
| RA | deg | Right ascension (J2000) |
| DEC | deg | Declination (J2000) |
| X | pixels | X position on detector |
| Y | pixels | Y position on detector |
| MAG_K | mag | K magnitude |
| ERROR_K | mag | K uncertainty |
| K_FLAG | | K flag (See Table 3.) |
| ERROR_TOT_K | mag | K total uncertainty |
| COL_J_K | mag | J−K color |
| ERROR_J_K | mag | J−K uncertainty |
| J_K_FLAG | | J−K flag (See Table 3.) |
| ERROR_TOT_J_K | mag | J−K total uncertainty |
| COL_H_K | mag | H−K color |
| ERROR_H_K | mag | H−K uncertainty |
| H_K_FLAG | | H−K flag (See Table 3.) |
| ERROR_TOT_H_K | mag | H−K total uncertainty |
| MAG_J | mag | J magnitude |
| ERROR_J | mag | J uncertainty |
| J_FLAG | | J flag (See Table 3.) |
| ERROR_TOT_J | mag | J total uncertainty |
| MAG_H | mag | H magnitude |
| ERROR_H | mag | H uncertainty |
| H_FLAG | | H flag (See Table 3.) |
| ERROR_TOT_H | mag | H total uncertainty |

To ensure consistency over our entire catalogue for each cluster, we then compared the magnitudes and colors in the overlap regions between each pawprint[2] and applied small (typically 0.01 to 0.02 mag) zero-point shifts to each pawprint to minimize the RMS of the differences in the overlap regions. As this procedure is carried out in each color as well as each magnitude, the smoothest change of color at array boundaries is provided by using the tabulated colors, not by differencing the tabulated magnitudes (which yield slightly different results because of the photometric calibration and normalisation procedures.) Our final catalogue presents magnitudes which are the means of all available measurements in the overlap regions.

## 7. THE DATA PRODUCTS.

Our primary data product is the catalog of sources within the area of the UKIRT surveys. Given the roughly 10 million objects in this catalog (see Col 7 of Table 1), this is available only as an electronic table, the column headers for which are presented in Table 2. There is no uncertainty quoted for the positions, but for stars with a signal-to-noise of 10 or better (in the $K$-band) we should use the error in astrometric calibration derived in Section 6, though formally this is an upper limit. For stars fainter than this, the precision of the positions of the stars measured on the array will be the dominant uncertainty, which we can estimate as the uncertainty in magnitude times the FWHM of the image (King 1983).

### 7.1. Object flagging

Each magnitude or color in the catalogue has a two character flag attached. For a colour (e.g. $J-K$) this is the flag for the $J$ and $K$ band measurements, whilst for single magnitudes the second character is the relevant flag, the first is always "O". The list of the flag meanings is given in Table 3, with the order in which they occur in the reduction process. This ordering is important as a flag described as strong will always be written, even if it overwrites another flag, whilst a weak flag will only be written if the current flag is "O".

[2] The area of sky observed by a single pointing of WFCAM is composed of four distinct regions separated by spaces nearly as large as each detector. This pattern is referred to as a pawprint.

### 7.2. The photometric uncertainties

Our final photometric uncertainties have several distinct components. In addition to those already discussed in Section 5.4, there is also an error in the photometric zero-point which derives from the profile correction. To measure this additional uncertainty we use the differences between the measurements of stars which lie in the overlaps between arrays. Since each pawprint has been placed on the 2MASS system, the correction we have to apply to bring the overlaps into agreement (see Section 6) gives an estimate of our profile correction error of between 1 and 2 percent (depending on crowding) in any single filter. Finally, when comparing stars with other photometric systems, there is the uncertainty in our photometric calibration. As explained in Hodgkin et al. (2009) this is better than 0.02 mags in fields where $E(B-V) < 2$. We therefore present two uncertainties in our tables, the basic heteroscedastic uncertainty from our estimate of the noise in each individual image, and then the combination of this (after allowing for pixel correlation) in quadrature with an uncertainty of 0.02 mags to allow for both the profile correction and the uncertainty in placing the photometry on a Vega system.

Crowding will also introduce errors in our photometry, the size of which we can estimate in an order-of-magnitude fashion. The number of stars in our fields roughly doubles in each successive magnitude bin, and if we assume this continues below our completeness limit we can calculate the magnitude at which the field becomes crowded. Our resolution element is roughly the area of the optimal extraction mask (typically $0.8''$ radius), which means the Eagle Nebula (our most dense field) becomes crowded at around 2.5 magnitudes below our 10$\sigma$ limit given in Table 1. Thus on average the photometry of a star at the 10$\sigma$ limit is affected by a star of roughly 10 percent of its flux, and so the likely crowding error is of order the quoted uncertainty. Using the fact that in flux space the uncertainties for each star are roughly the same as long as the star is faint compared with the sky (the "background limited case", e.g. Naylor 1998), it follows that over most of our magnitude range the likely perturbation for a star is of order an error bar. Conversely we can see that the contamination in the Rosette Nebula occurs at fluxes a factor 8 fainter, corresponding to only about 10 percent of an error bar. Dealing with the crowding in a statistically robust way lies beyond the scope of this paper, but this order-of-magnitude calculation indicates that for the MYStIX project, provided we use the optimal extraction, it is never a major issue. This contrasts with the situation for a $2''$ aperture, when the contamination would be a factor six worse, and be the dominating issue for about two-thirds of our fields.

### 7.3. False Positives and False Negatives

**Table 3**
Flags for the UKIRT photometric catalog.

| Flag | Meaning | algorithm | Power |
|---|---|---|---|
| O | O.K. | | Weak |
| S | Saturated pixel | Pixel flagging | Strong |
| F | Bad pixel | Pixel flagging | Strong |
| N | Non-stellar | Star shape estimator | Strong |
| B | Background fit failed | Sky determination | Weak |
| M | Negative (Minus) counts[a] | Flux measurement | Weak |
| V | Variable | Combining measurements | Weak |

[a] The measured flux is negative, the magnitude should be used as a bright limit.

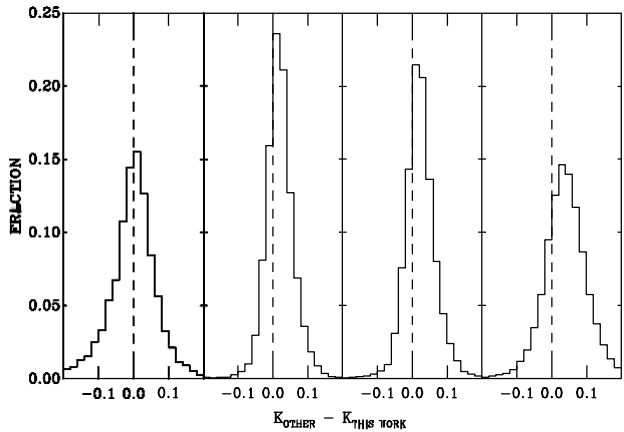

**Figure 3.** A comparison of the magnitude differences between the reduction presented here and (from left to right) 2MASS and the UKIDSS $1''$ aperture photometry for $12<K<14$, $14<K<16$ and $16.5<K<17.5$.

Visual examination of a field with high star density (low Galactic longitude) and high nebulosity reveals a number of errors in our catalog. False positives are present in several circumstances: very bright stars (spurious sources within saturated stellar profiles and along diffraction spikes - see also Solin et al. 2012), intermediate brightness stars (spurious sources on edge of profiles of bright objects), faint nebulosity (clusters of spurious sources), edges of bright rimmed clouds (line of spurious sources due to rapid gradient in surface brightness), and associated with closely spaced stars (spurious source introduced between two stars separated by $\sim 2''$). False negatives include some missed stars $1''$ from bright stars, missed faint stars in dense obscuring clouds with rapid surface brightness gradients, and missed stars in bright filamentary nebulosity. Some of these problems are illustrated in Figure 5 of Feigelson et al. (2013). In the most crowded and nebular MYStIX fields, these false positives and negatives are estimated to be present at the level of $\sim 2\%$ of the source catalog.

## 8. COMPARISON WITH THE STANDARD DATA PRODUCTS

Figure 3 shows the histograms comparing our photometry with 2MASS (left) and bright, medium and faint magnitude slices of the UKIDSS data for the Rosette Nebula field (chosen to avoid the effects of crowding). The data show excellent agreement, with the zero points agreeing to within about a percent and any shift with magnitude between the bright and faint UKIRT data being around a percent.

The small extra shift in the rightmost (faintest star) histogram is due to a well-known effect of using a small aperture; the centroid is dragged towards the brightest part of the image by noise, placing "too much" flux in the small aperture (or in our case the optimal extraction mask, which has a smaller effective diameter than $1''$). The same effect is present in the UKIDSS data alone. Comparing the UKIDSS $1''$ and $2''$ apertures gives a mean difference of 2 percent, in the sense that the smaller aperture data are brighter. Comparing the optimal extraction with the UKIDSS $2''$ aperture measures the total bias as 3 percent between $16.5<K<17.5$ (equivalently 0.06 to 0.10 mags in uncertainty in the optimal extraction), which is close to values we obtained from simulation. Although this shift should not be present in $J$ and $H$, since our positions are taken from the $K$-band image, a smaller, unexplained shift does remain, but the optimal and UKIDSS $1''$ aperture reductions still agree at the roughly one percent level down to a signal-to-noise of 10.

The crucial point here is that small apertures or optimal extractions deliver better signal-to-noise and are more robust against crowding, but at the cost of a small shift to brighter magnitudes at the faint limit. Which to use is dependent on the science being undertaken. In our case, where over 90 percent of the sources identified as young stars have uncertainties less than 0.02 mags, it is clear that the advantage of a factor six decrease in the crowding for the optimal extraction (see Section 7.2) far outweighs a bias at faint magnitudes.

There are slightly larger differences at faint magnitudes in the most crowded fields. The $1''$ aperture data are five percent fainter than the optimal extraction at the completeness limit of $K=16$ in the Trifid Nebula (one of our highest density fields). This is due to the difficulty of defining the background in this crowded field. To demonstrate this, we determined the optimally extracted flux at a set of 3000 randomly chosen positions in the Trifid field. The modal flux at these postions is zero, implying our background subtraction has worked well. However, the median flux is brighter by roughly the difference between the optimal extraction and the UKIDSS $1''$ aperture. This re-emphasises the fact that crowding affects photometry in the most dense fields, where even the definition of the background is unclear, but as pointed out in Section 7.2 it is not at a level which is important for MYStIX.

The most obvious difference between the standard UKIDSS data products and the catalogues presented here is that in the optimal reduction the uncertainties for a given star are smaller. Taking DR 21 as a field of typical stellar density, we can compare stars in the range $16.5 < K < 17.5$, since at the end of this range the quoted uncertainty in the standard data product is around 0.1 mags, and so the catalog should be substantially complete. In this range we find that the uncertainties in our reduction are approximately 0.9 times smaller than those of the standard method. The data for Figure 1 suggest that for a seeing of $1''$, the ratio of the signal-to-noise in a $1''$ aperture to an optimal extraction should indeed be about 0.85. At brighter magnitudes the uncertainties become dominated by systematics, and we should compare the 0.02 mags night-to-night repeatability found by Hodgkin et al. (2009) with our own estimate of the profile correction error of 1 or 2 percent.

The number of detected sources is significantly larger in our reduction than in the standard one. For a large area in DR 21 we compared the number of stars in the standard data products which are unlikely to be noise or galaxies (i.e. with MergedClass neither 0 nor $-3$) and have a $K$-band uncertainty of less than 0.3 mags with the number of stars in our catalog with unflagged $K$-band magnitudes and uncertainties $< 0.27$ mags (the latter figure allowing for the difference in photometric precision). We find there are about 75 percent more objects in our reduction, which mainly fall in the range $K = 18 - 19$, when the uncertainly in our catalogue declines below 0.15 mags. The differences are less extreme in the more crowded case of the Trifid Nebula, but even here the optimal reduction has around 20 percent more sources, again concentrated close to the completeness limit.

## 9. THE COLOR-MAGNITUDE AND COLOR-COLOR DIAGRAMS.

Figure 4 shows the color-magnitude and color-color diagrams for the regions of sky covered by the X-ray observations in each of the MYStIX fields for which we have UKIRT data. There is a large diversity in the morphology of these diagrams for both the field stars and the probable members of the star-forming complexes (MYStIX Probable Complex Members - MPCMs) identified by Broos et al. (2013).

### 9.1. *Theoretical color-magnitude diagrams*

To better understand this diversity we show in Figure 5 a theoretical 5 Myr old population in the $K$ vs $J - K$ color-magnitude diagram (CMD) and $J - H$ vs $H - K$ color-color diagram. The color-magnitude diagram makes clear the relatively narrow range in the color of the stars, with the isochrone being almost vertical for $M_J < 0$ or $M_J > 3$. For hot stars this is because the $J$ and $K$ filters lie in the Rayleigh-Jeans tail of the spectral energy distribution. For cool stars, the bend toward the vertical occurs at $T_{\rm eff} \simeq 4\,000$K, where a blackbody would continue redwards. This turnover is due to flux redistribution by $H_2O$ opacity into the $H$ and $K$ bands from the wavelength regions between and to either side of these bands, since $H$ and $K$ filters are placed in regions where the water opacity is low (Allard & Hauschildt 1995). This virtual independence of the color on effective temperature, especially the region below $T_{\rm eff} \simeq 4\,000$K means that position along the color axis of the CMD is primarily driven by extinction (with a small effect due to disks for pre-main-sequence stars). Thus the morphology of the field stars in the CMD is driven by a combination of the populations visible along the line of sight through the star-forming region, and the relationship between extinction and distance.

### 9.2. *Theoretical color-color diagrams*

Like the color-magnitude diagram, the theoretical color-color diagram turns over at 4000 K, corresponding to $J - H$=0.7, $H - K$=0.2 for a 5 Myr sequence (the right-hand panel of Figure 5). Whilst we expect the majority of stars to fall either along this sequence or along the reddening vectors at its extremes (marked as lines in the right-hand panel of Figure 5), there are stars which can fall outside this regime. We have illustrated this by adding the colors of all stars hotter than 2000 K in the BTSettl atmosphere library (Allard et al. 2011) as red dots. There is a cool tail of stars between 2000 and 3000 K which lie below the lower reddening vector, and a group of warmer, but low-gravity atmospheres (corresponding to giants) which lie above the 5 Myr sequence, some of which are above the reddening vector.

### 9.3. *MYStIX color-magnitude diagrams*

The order we have chosen to display the colour-magnitude and color-colour diagrams in Figure 4 is roughly that of increasing numbers of red stars. This ordering correlates primarily with Galactic longitude, with fields closer to the Galactic center having larger numbers of reddened field stars along the line of sight.

Beginning with the fields more than $90°$ from the Galactic Center, NGC 2362 is a particularly simple CMD, being composed of two sequences at different extinctions; a red sequence which coincides with the young stars, and a lower extinction sequence of field stars. Blueward of $J-K$=1.5 we see similar morphologies for NGC 1893, NGC 2264 and the Rosette, though NGC 2264 lacks the nearby field sequence. However these clusters show an increasing number of objects redward of $J-K = 1.5$, reflecting an increasing number of background stars.

The remaining fields are within $20°$ of the Galactic Center, excepting DR21 with $l$=82. Both DR 21 and the Lagoon Nebula CMDs have a significant number of stars redward of $J-K$=1.5, though the morphology of the red part of the CMD has no clear structure. The fields with the largest numbers of stars redward of $J-K$=1.5 show more structured red field-star CMD morphologies. Thus the Eagle Nebula and M 17 have well developed giant sequences running redwards and fainter from $J-K \approx 2.5$ $K \approx 13$, whilst in NGC 6334, NGC 6357 and the Trifid Nebula the stars redward of $J-K$=1.5 are bimodal in color distribution. As explained in Lucas et al. (2008), the details of the red part of the CMD depend on whether one is sampling the disk or bulge giants, which in turn depends on the line of sight, and, of course, its associated extinction.

In the early regions in our ordering (NGC2362 to the Rosette Nebula) the young stars largely lie in a single almost vertical sequence, implying a relatively uniform extinction to the cluster, although there are a small fraction of objects which are clearly much redder. In contrast, the objects from the first and fourth Galactic quadrants have young stars which are more scattered in colour, varying from DR21 with no discernible sequence, to the Lagoon Nebula which has the tightest sequence in this group. For all regions the red colors of most stars are primarily an extinction effect; it is not caused by a $K$-band excess due to disks, as the objects have similar scatter in the $J$ vs $J-H$ CMD. The morphology of the background stars in the CMD is no guide to how the pre-main-sequence stars will be distributed; in the case of the Eagle the MPCMs are restricted to a relatively narrow range of color, whilst M 17, where the CMD is otherwise similar, has members stretched over the entire color range. This decoupling between field star and member extinctions is what we might expect, given sight lines to field stars that sample both the extinction of the star-forming complex, and interstellar medium beyond it.

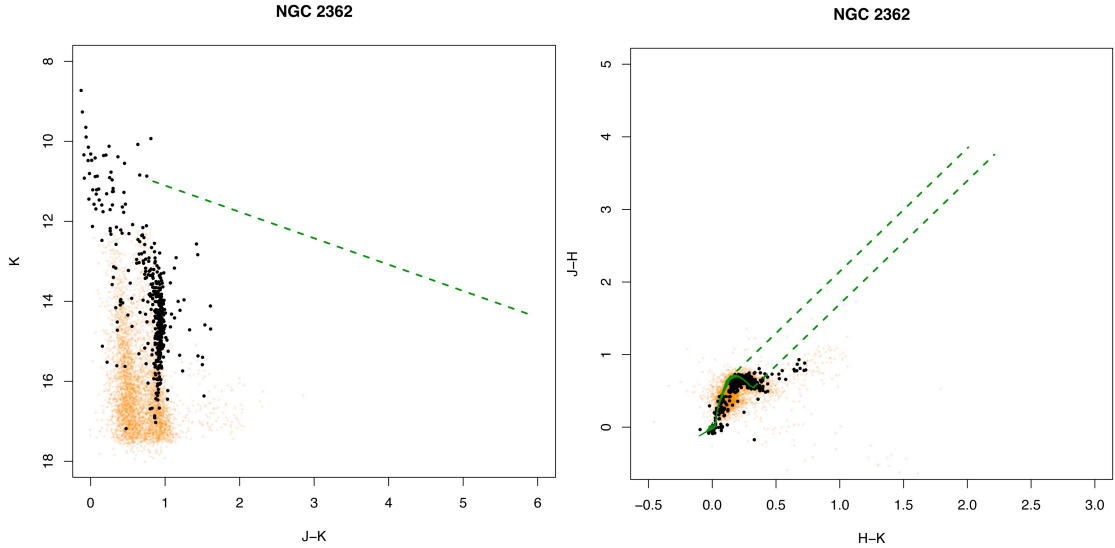

**Figure 4.** The color-magnitude and color-color diagrams for each star-forming complex for the region of sky covered by the X-ray observations. The black dots are the MYStIX complex probable members, and the orange dots other stars in the field of view. Only stars with uncertainties of less than 0.1 mag in all three filters and in $J-K$ and $H-K$ and unflagged in all three filters are shown. The green curve is the same 5 Myr isochrone presented in Figure 5 and the green dashed lines $A_V = 30$ extinction vectors from Rieke & Lebofsky (1985). (The diagrams for NGC 2362 are shown as a sample in the printed edition, the remaining regions appearing only in the electronic edition.)

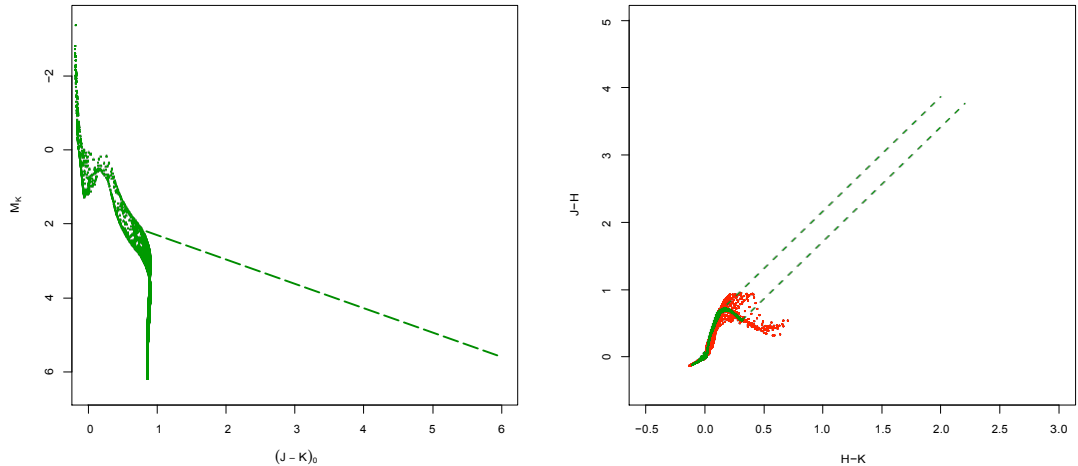

**Figure 5.** Theoretical color-magnitude and color-color diagrams for a 5 Myr-old isochronal population (green dots). Below $5M_\odot$ the isochrones are from Dotter et al. (2008) converted into the color-magnitude plane using the Phoenix BT-Settl model atmospheres (Allard et al. 2011), folded through the filter responses given by Hewett et al. (2006). Above $5M_\odot$ the interior models are those of Schaller et al. (1992) with the atmospheres of Castelli & Kurucz (2004). The distribution with magnitude is derived using a Kroupa mass function (Dabringhausen et al. 2008) and binarity is dealt with as described in Naylor (2009) and Naylor & Mayne (2010). The red dots in the color-color diagram are the colors of all atmospheres hotter than 2 000K from Allard et al. (2011), the green dotted lines are extinction vectors from Rieke & Lebofsky (1985) corresponding to $A_V = 30$ magnitudes.

### 9.4. *MYStIX color-color diagrams*

Examining the color-color diagrams in the order we used for the color-magnitude diagrams, we see an increasing spread of the field stars along the color-color extinction line. This is a natural consequence of the order that was chosen of increasing numbers of red objects in the CMDs. We also see that in the fields far from the Galactic Center there are few stars in the upper half of the region between our reddening vectors, in contrast to the fields closer to the Galactic Center. This region of the color-color diagram is normally associated with reddened giants (with the lower half of the region between the vectors occupied by dwarfs). Whilst Figure 5 shows main-sequence stars can be reddened into the upper region, the paucity of stars in this part of the CMD in the second and third Galactic quadrants is undoubtedly due to the low numbers of giants.

A significant number of stars lie below the reddening lines. In the case of the field stars most of these are probably caused by crowding, the colors resulting from the combination of a blue and a red star. For the members, however, this is primarily the effect of the $K$-band excess caused by disks.

## 10. CONCLUSIONS

We have presented the data reduction process for the deep near-IR UKIRT data used in the MYStIX project. In doing so we have demonstrated that optimal extraction can work well in relatively crowded fields, provided certain adjustments are made to the extraction parameters normally used. Compared with the standard UKIDSS data products, the photometry developed here is more robust against crowding, and contains more sources.

We are grateful to Tom Kerr and Gary Davis for an allocation of Director's Discretionary Time which changed this part of the MYStIX project from being useful extra data from a few clusters which happened to be in the GPS, to a dataset which covered the majority of our clusters, and so became central to our analysis. The United Kingdom Infrared Telescope is operated by the Joint Astronomy Center on behalf of the Science and Technology Facilities Council of the U.K. This work is based in part on data obtained as part of the UKIRT Infrared Deep Sky Survey and in part on data obtained in UKIRT director's discretionary time.

We are also grateful to Charles Williams for his support of the Apple Mac X-grid cluster in Exeter, a facility which makes it possible to reduce the large amounts of data this part of the MYStIX project uses.

The MYStIX project is supported at Penn State by NASA grant NNX09AC74G, NSF grant AST-0908038, and the Chandra ACIS Team contract SV4-74018 (G. Garmire & L. Townsley, Principal Investigators), issued by the Chandra X-ray Center, which is operated by the Smithsonian Astrophysical Observatory for and on behalf of NASA under contract NAS8-03060. This research made use of data products from the *Chandra* Data Archive.

This publication makes use of data products from the Two Micron All Sky Survey, which is a joint project of the University of Massachusetts and the Infrared Processing and Analysis Center/California Institute of Technology, funded by the National Aeronautics and Space Administration and the National Science Foundation.

Facilities: UKIRT (WFCAM)